\documentclass{aa}  

\usepackage{graphicx}
\usepackage{txfonts}
\usepackage{xcolor}
\def\kms{{\rm km\, s^{-1}}}

\begin{document} 

   \titlerunning{Barred AGN galaxies in paired systems: exploring the impact on nuclear activity}
   \authorrunning{Alonso et al.}

   \title{Barred AGN galaxies in paired systems: exploring the impact on nuclear activity}


   \author{Sol Alonso\inst{1}, 
          Matias Vera-Rueda\inst{1}, 
          Georgina Coldwell\inst{1},
          Fernanda Duplancic\inst{1}
          \and
          Valeria Mesa\inst{2, 3 , 4}
          }

   \institute{Departamento de Geof\'{i}sica y Astronom\'{i}a, CONICET, Facultad de Ciencias Exactas, F\'{i}sicas y Naturales, Universidad Nacional de San Juan, Av. Ignacio de la Roza 590 (O), J5402DCS, Rivadavia, San Juan, Argentina,\\         \email{solalonsog@gmail.com}
         \and
            Instituto de Investigación Multidisciplinar en Ciencia y Tecnología, Universidad de La Serena, Raúl Bitrán 1305, La Serena, Chile.
         \and Association of Universities for Research in Astronomy (AURA)
\and 
Grupo de Astrofísica Extragaláctica-IANIGLA, CONICET, Universidad Nacional de Cuyo (UNCuyo), Gobierno de Mendoza
             }

   \date{Received xxx; accepted xxx}

 
  \abstract
   {}
   {To unveil the influence of galaxy-galaxy interactions on the material transport driven by galactic bars towards the central regions of the active galactic nuclei (AGN) galaxies, and to assess the efficiency of the combined mechanisms of interactions and bars in fueling massive black holes, we meticulously examine barred active  galaxies in paired systems.
}
   {Our study focuses on barred AGN galaxies in pairs with 
 projected separations of $r_{p}< 100 \rm \,kpc \,h^{-1}$
and relative radial velocities of $\Delta V< 500 \rm \,km \,s^{-1}$ within $z<0.1$,  identified from Sloan Digital Sky Survey (SDSS). To quantify the impact of interactions on material transport by galactic bars, we also constructed a suitable control sample of barred active  galaxies without paired companions, matched in redshift, absolute r-band magnitude, stellar mass, color, and stellar age distributions. 
Additionally, we calculated the structural characteristics of galactic bars through two-dimensional image modeling, considering that bars exhibit a wide range of shapes and sizes, which may influence their ability to channel material.
}
   {From this study, we clearly found that nuclear activity (as derived from the $Lum[OIII]$) increases as the projected separations between galaxy pair members decrease. Notably, barred AGN galaxies in close pairs ($r_p \lessapprox$ 25 kpc $h^{-1}$) exhibit significantly higher nuclear activity compared to galaxies in the control sample. 
Additionally, barred galaxies with a close pair companion show enhanced nuclear activity across all ranges of luminosity, stellar mass, and color.
We also found that barred AGN galaxies with longer bar structures exhibit more efficient nuclear activity compared to those with shorter bars. This trend is especially pronounced in barred AGN galaxies within close pair systems, which show a significant excess of high $Lum[OIII]$ values.
Furthermore, we examined the central nuclear activity in barred AGNs undergoing major and minor interactions. Our findings show a clear escalation in nuclear activity as the pair projected separations decrease, particularly pronounced in major systems. Additionally, nuclear activity distributions in barred AGN samples within major and minor pairs exhibit similar trends. However, a significant deviation occurs among barred AGN galaxies in close pair systems within major interactions, showing a substantial excess of high $Lum[OIII]$ values. 
This result is also reflected in the analysis of the accretion strength onto central black holes.
These findings indicate that external perturbations from a nearby galaxy companion can influence gas flows induced by galactic bars, leading to increased nuclear activity in barred AGN galaxies within pair systems. Thus, the coexistence of both — bars and interactions — significantly amplifies central nuclear activity, thereby influencing the accretion processes onto massive black holes.
}
   {}

   \keywords{galaxies: active - galaxies: interactions
- barred galaxies
               }

   \maketitle
%

\section{Introduction}

The accretion of material onto a central massive black hole is widely accepted as the origin of active galactic nuclei (AGN). This process is believed to trigger nuclear activity and drive the growth of supermassive black holes \citep{Lyn69, Rees84}. Consequently, dynamic perturbations are proposed as the primary mechanisms driving the accretion process and the subsequent growth of supermassive black holes. In this context, numerous authors agree that galactic bars and galaxy mergers/interactions are typically regarded as the two dominant processes responsible for the accumulation of material in the centers of active galaxies 
\citep[e.g.][]{Combes93, Miho96, Alonso2007, Alonso2013, Alonso2014, Alonso2018, stor01, saty14, ellison15, her16, stor19, choi24}.

In the local universe, galactic bars are observed in a significant fraction of spiral galaxies, and their presence is influenced by the properties of the host galaxies \citep{Master10, oh12, sanchez14, james16, Vera, Lee19, lin20, zee23}. These structures are believed to play a crucial role in the dynamical evolution of their hosts \citep{ann2000, Ellison2011, zhou15, gero21, gero24, yu22}. Numerous simulations have demonstrated that bars can efficiently transport gas from the outer regions to the innermost central areas of barred galaxies \citep{DeBat98, Atha03, Petersen19}. The interaction of gas clouds with the edges of the bars results in the generation of shocks, leading to angular momentum loss and driving the flow of material toward the central kiloparsec scale \citep{Shlo90}.

It has been demonstrated that the gas infall produced by bars toward the innermost regions of galaxies is an efficient mechanism for triggering nuclear activity in the central zones of active galaxies \citep{Combes93, combes2000, Jung18}. In their work, \cite{Alonso2013} used Sloan Digital Sky Survey \citep[SDSS,][]{York} data to study isolated barred AGN galaxies (AGNs) finding a higher prevalence of powerful nuclear activity compared to a suitable control sample of unbarred AGNs with similar distributions of redshift, magnitude, morphology and local environment. Additionally, they observed that barred AGNs  have an excess of objects with high accretion rate values compared to unbarred ones.

Furthermore, the investigation carried out by \cite{Alonso2014} on barred AGN spiral galaxies within groups and clusters revealed that the enhancement of nuclear activity induced by bar perturbations is also evident among barred active galaxies in high-density environments.
Following this, \cite{Gallo15} identified a higher prevalence of barred AGNs compared to star--forming barred galaxies. However, they found no correlation between the accretion rate of the central black hole and the presence of a bar.  \cite{Cheung15} extended the study to higher redshifts, concluding that large-scale bars cannot be considered the primary fueling mechanism for the growth of supermassive black holes, in agreement with the findings of \cite{Gould17}.

In more recent studies, \cite{galrland23} analyzed a sample of merger-free active galactic nuclei taken from \cite{simmons17}. Their findings indicated that AGNs are marginally more likely to host a bar than inactive galaxies. This suggests that bars may potentially trigger AGN activity.
Also, \cite{Kata24} used the IllustrisTNG100 cosmological magnetohydrodynamical simulations to investigate the distribution of black hole masses in barred and unbarred galaxies. Their results showed that the median black hole mass in barred galaxies is higher than that in unbarred ones, suggesting a significant influence of stellar mass on black hole growth.

Conversely, observational evidence suggests a correlation between mergers/interactions and the feeding of central black holes. In a comprehensive statistical analysis of nuclear activity, \cite{Alonso2007} examined AGNs in 1607 close pairs as well as AGNs without companions. Their findings indicated that active galaxies with signs of strong interaction exhibit significantly greater nuclear activity than isolated AGNs. The accretion rate also indicates that AGN in merging pairs are actively feeding their central black holes. Further supporting this, \cite{Mesa2014} analyzed a sample of spiral galaxy pairs from the SDSS, classifying them based on the rotation patterns of their spiral arms. The results demonstrated that systems with spiral arms rotating in opposite directions exhibit increased nuclear activity.

Moreover, \cite{duplancic21} performed a comparative study on AGN galaxies in small systems, such as pairs, triplets, and groups. They used BPT \citep{BPT} and WHAN \citep{cidf11} diagnostic diagrams to identify optical AGNs and also include  WISE data to detect mid-IR AGNs \citep{assef18}. 
Their studies highlight the significant role of interactions, beside the influence of the global environment, in the activation of the AGN phenomenon in small galaxy systems.

In addition, bars can themselves be the result of galaxy-galaxy interactions, as evidenced by studies such as those from \cite{nogu87,kaza08,moe17}. Consequently, during close encounters between galaxies, there is a significant redistribution of mass and a strong gravitational tidal torque that can result in the formation of a central bar. Therefore, it can be suggested that close encounters between galaxies lead to substantial mass redistribution and strong gravitational tidal forces, potentially forming a central bar.
Furthermore, \cite{Cava20} employed N-body simulations to identify a robust correlation between mergers and bar formation. These authors delineated two distinct phases in this process: the initial, tidally induced formation preceding the merger; and the destruction and/or regeneration of the bar during and following the merger. 

A compelling analysis by \citet[][hereafter A18]{Alonso2018} explores the role of bars versus galaxy interactions in central black hole feeding. This study investigates the influence of internal processes within bars compared to the external mechanisms of galaxy interactions on central nuclear activity. Using a large and homogeneous sample of barred AGN galaxies and AGNs in pair systems from the SDSS, the study shows that both mechanisms play an important role in driving radial gas inflows towards the innermost region of galaxies. This leads to an enhancement in nuclear activity and the accretion rate of central black holes in spiral active nuclei galaxies. It is also worth noting that the internal process of the bar perturbation is more effective at transporting the gas flow to the central zones than the external mechanism of mergers and interactions.

Previous research has independently investigated the mechanisms of AGN activity enhancement through the presence of bars and galaxy interactions. Gaining insight into the combined effects of these processes on supermassive black holes will deepen our understanding of their feeding dynamics allowing  us to make further advances in this field. 
In the present work, we analyze a sample of barred AGN galaxies in paired systems to examine the impact on nuclear activity.

The structure of the paper is organized as follows: Section 2 describes the samples under study, including the construction of the control sample and the estimation of bar parameters. In Section 3, we conduct a detailed analysis of the effects of interactions in barred AGN galaxies, focusing on how these interactions influence nuclear activity. 
Section 4 examines the relationship between AGN activity and bar strength parameters.
Section 5 provides a comprehensive study of barred AGN in both major and minor galaxy pairs, examining the differences in nuclear activity and analyzing the accretion strength of central black holes. Finally, Section 6 presents our main conclusions, summarizing the key findings and their implications for understanding black hole activity triggered by galaxy interactions in barred galaxies. Throughout this paper, we have assumed a $\Lambda$-dominated cosmology, 
with $\Omega_{m} = 0.3$, $\Omega_{\lambda} = 0.7$ and $H_0 = 100 \kms Mpc^{-1}$.

\section{Barred AGN galaxies in pair systems}

\begin{figure*}
 \centering
 \includegraphics[width=1.\textwidth]{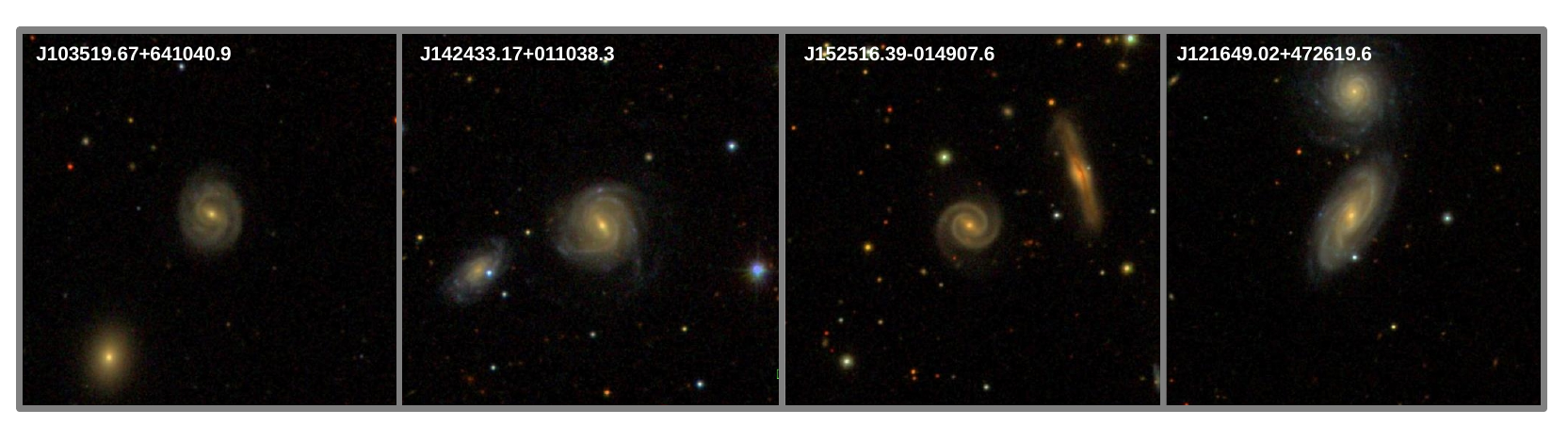}
\caption{Images of typical examples of barred AGN galaxies in pair systems obtained from our sample. The scale of the images is $\approx$ 2$'$.5 x 2$'$.5. 
 }
  \label{ej}
\end{figure*}

In this study, we used a subset of AGN in barred galaxies derived from A18. This sample was extracted from the SDSS Data Release 7 \citep{Abaza2009}, which comprises 11663 square degrees of sky imaged in five band widths (u, g, r, i, and z). The main galaxy sample of the SDSS is limited to $r$-band magnitude of $r_{lim}<17.77$, covering a redshift range of 0 < z < 0.25, with a median redshift of 0.1 \citep{Strauss2002}.
In A18, we also included several galaxy properties from the  MPA/JHU%
\footnote{http://www.mpa-garching.mpg.de/SDSS/DR7/ %
} and the NYU%
\footnote{http://sdss.physics.nyu.edu/vagc/ %
} catalogs, such as stellar mass content, indicators of star-bursts, emission-line fluxes, etc. \citep{Brinchmann2004, Blanton2005}.

The methodology for detecting barred AGN galaxies within paired systems is based on A18. Below is a condensed summary of the key elements used to compile the sample for the present work:

\begin{itemize}
\item Active galaxies were identified on SDSS data using a standard diagnostic diagram proposed by \cite{BPT}, applying the criteria established by \cite{Kauffmann2003} and using publicly MPA/JHU available emission-line fluxes.
    
\item To select spiral hosts of active galaxies, a cross-correlation was conducted between AGN galaxies and spiral objects identified through Galaxy Zoo\textbf{1}\footnote{http://www.galaxyzoo.org/} \citep{Lintott2008,Lintott2011}. 

\item In order to identify bar-like structures within spirals, a further criteria were applied aiming to facilitate a classification based on the naked-eye detection. We restrict AGN spirals to redshifts of z < 0.1, g-band magnitudes brighter than 16.5, and axial ratios (b/a) > 0.4.
This process yielded a sample of 6772 face-on bright AGN spiral galaxies.

\item Within this sample, a visual inspection was made using combined $g$ + $r$ + $i$ color images from the online SDSS Image Tool%
\footnote{http://skyserver.sdss.org/dr7/en/tools/chart/list.asp %
}, identifying 1927 AGNs hosted in barred galaxies, constituting approximately 28.5$\%$.

\item Barred active objects in paired systems were identified by cross-correlating the total barred AGN sample with an SDSS pair catalog containing 2970 AGN spiral galaxies in pairs. The selected double systems met specific criteria: galaxies with a g-band magnitude brighter than 16.5, projected separations between members ($r_p$) less than 100 kpc $h^{-1}$, and relative radial velocities ($\Delta$V) less than 500 km $s^{-1}$, all within a redshift range of z < 0.1. This process yielded a subset of 446 barred AGN galaxies within paired systems.
The remaining 1481 barred AGN galaxies do not have nearby companions.

\end{itemize}

These findings reveal that 23.1$\%$ of the total 1927 barred AGN galaxies are located within such pair configuration. 
Additionally, we compared this with the fraction of non-barred AGN in pair systems, which is 19.4$\%$. 
This  indicates a higher proportion of barred AGN compared to unbarred AGN in similar paired system configurations.
This suggests that galaxy interactions can stimulate bar formation, although they are not the only process involved. 
In this direction, several authors have shown that bar structures may emerge as a consequence of galactic interactions \citep[e.g.][]{moe17,nogu87}.
\cite{kaza08} demonstrated that satellite galaxies can significantly impact the timing of bar formation, suggesting that encounters indirectly induce perturbations across the disc, which evolve into delayed waves in the central region and interact with an emerging seed bar. Additionally,  \cite{Ski12} found that bars can also form through secular evolution, highlighting a dichotomy between internal secular mechanisms and external effects of neighboring galaxies.

Fig. \ref{ej} shows images depicting typical examples of barred AGN galaxies within pair systems selected from our sample.

\subsection{Control Sample}

To investigate how galactic interactions impact the central nuclear activity of black holes in barred AGN galaxies, we used the sample of 1481 barred AGN galaxies that lack nearby companions within specified projected separation and velocity criteria ($r_p$ < 100 kpc $h^{-1}$ and $\Delta$V < 500 km $s^{-1}$). From this dataset, we constructed a suitable control sample using a Monte Carlo algorithm to select galaxies that match the redshift, r-band absolute magnitude, and stellar mass distributions of the barred AGN galaxies in pairs sample \citep{Perez_2009}.

\begin{figure}
 \centering
 \includegraphics[width=.4\textwidth]{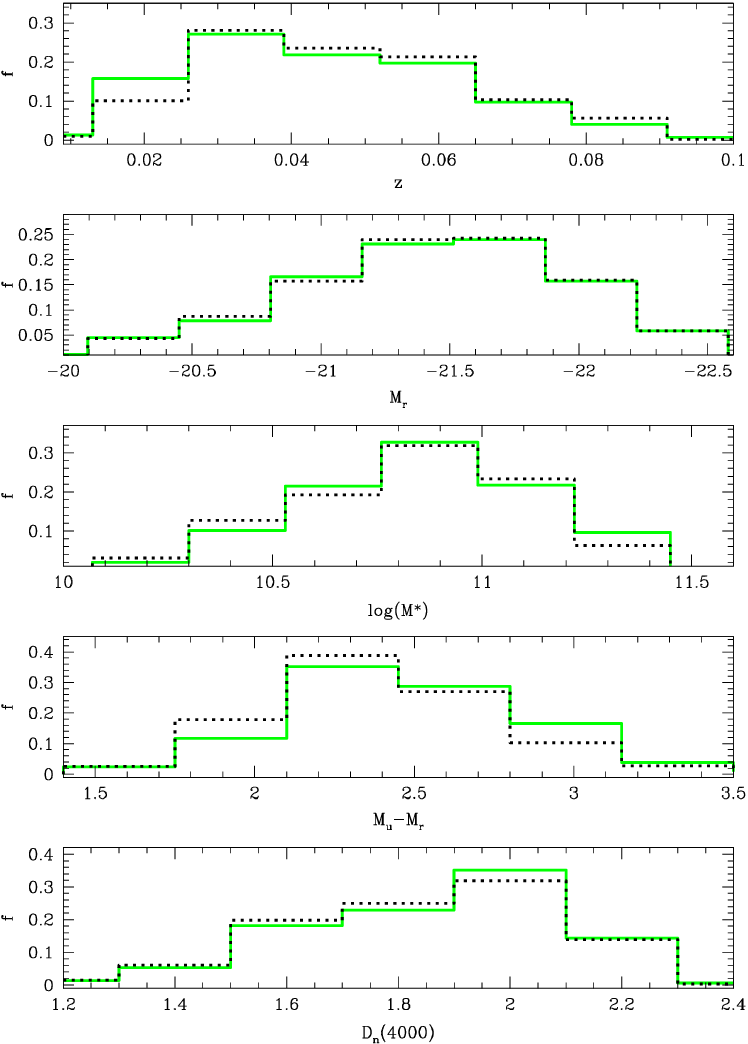}
\caption{Normalized distributions of redshift, $z$,  r-band absolute magnitude, $M_r$, stellar mass, $log(M_*)$, color, $M_u - M_r$ and age of stellar population parameter, $D_n(4000)$, 
for barred AGN galaxies in pairs (solid lines) and barred AGNs in the control sample (dotted lines).
 }
  \label{control}
\end{figure}

Additionally, we required the control galaxies to have comparable host characteristics to those of the barred AGNs in pairs. To ensure a rigorous analysis, we selected control objects that exhibit similar distributions in color and stellar age population as the barred AGN galaxies within paired systems.
In the context of stellar population parameters, we have adopted the spectral $D_n(4000)$ index as a key indicator.
This index holds significance as it reflects the spectral features of aging stars, particularly those occurring around 4000$\AA$ \citep{Kauffmann2003}. This phenomenon arises from the accumulation of numerous spectral lines within a narrow spectrum range. In this study, we have adopted the definition outlined by \cite{balo99}.
In Fig. \ref{control}, we present the distributions of z, $M_r$, $log(M_*)$, $M_u - M_r$, and $D_n(4000)$ for the barred AGN pair galaxy catalog (solid lines) as well as the corresponding control sample (dashed lines).

We conducted Kolmogorov-Smirnov (KS) tests between the distributions of the control sample and those of barred AGN galaxies within pair systems. This yielded a $p$-value indicating the probability that a value of the KS statistic would be as extreme as or more extreme than the observed value if the null hypothesis were true. In all instances, we obtained $p$ $>$ 0.05, supporting the null hypothesis that the samples were drawn from the same distributions.
Finally, the control sample comprises 446 barred AGN galaxies without paired companion.

In this context, the methodology used in the construction of these catalogs guarantees that the samples share similar host properties. Consequently, they can be used to unveil the efficiency of the coexistence of both mechanisms, interactions + bars, compared to the exclusive effect of galactic bars on central nuclear activity. Additionally, this analysis will help reveal the impact of galaxy interactions on the dynamics of gas transported by the galactic bars.

\subsection{Bar parameters and measurements}

\begin{figure*}
 \centering
 \includegraphics[width=0.8\textwidth]{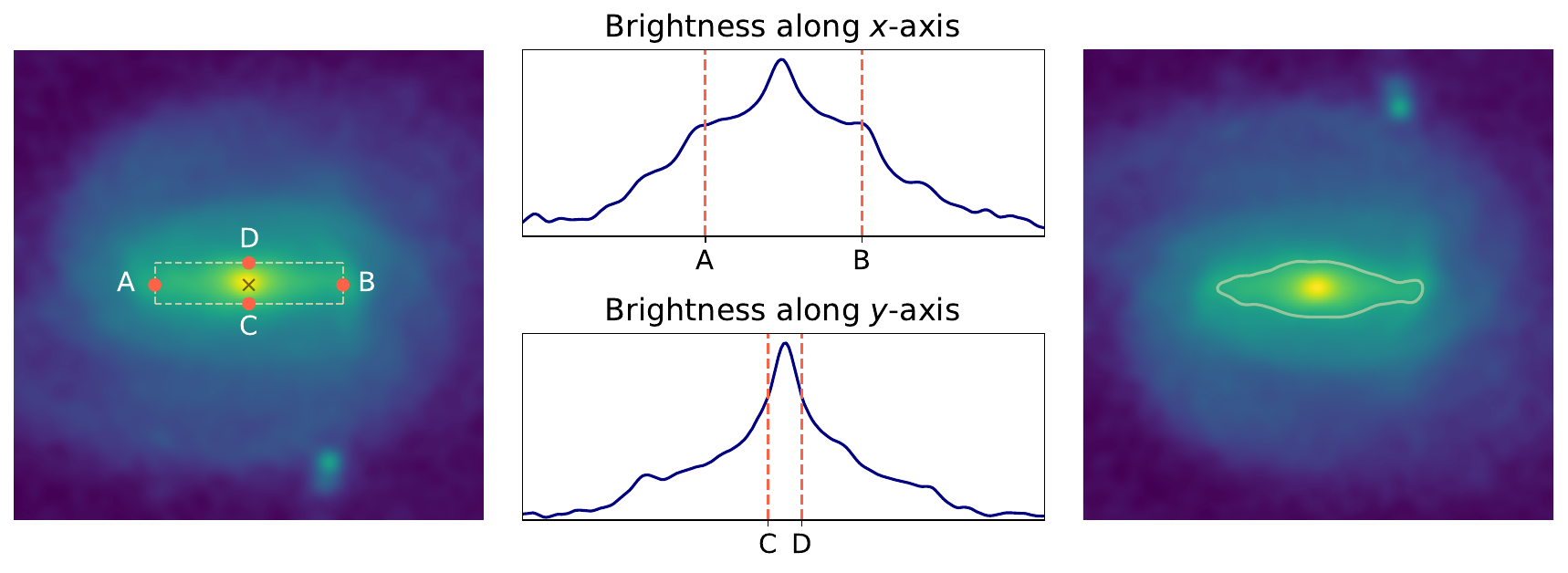}
\caption{Schematic representation of the procedure used to obtain an iso-brightness contour enclosing the bar, based on the average of brightness measurements from four manually selected points, two lying on the $x$ and two on the $y$ axes of the image.
 }
  \label{BarBrightness}
\end{figure*}

\begin{figure}
 \centering
 \includegraphics[width=.45\textwidth]{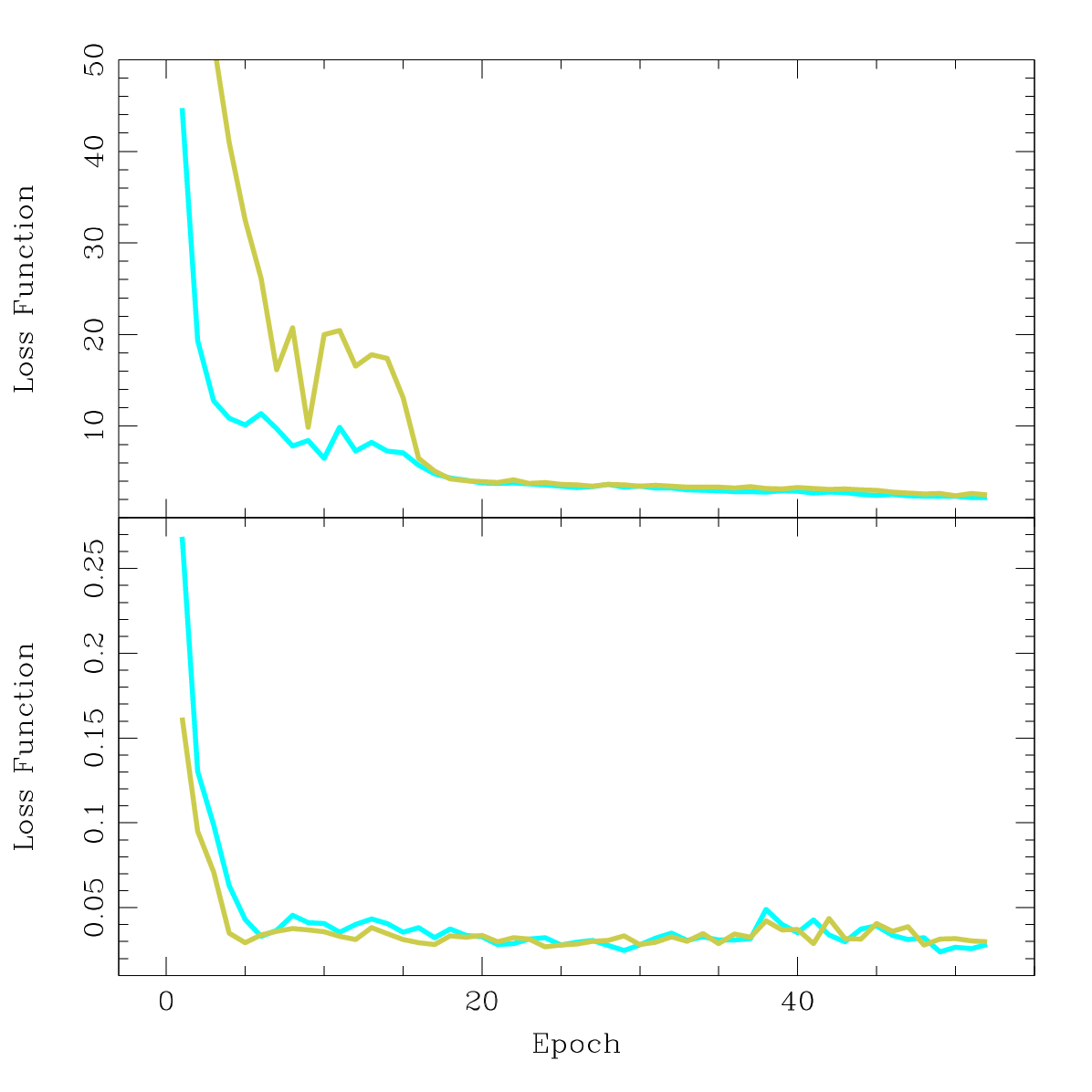}
\caption{Evolution of the loss function over epochs for the training (orange lines) and validation (cyan lines) sets, shown for the convolutional neural network used for image rotation (top panel) and the standard neural network used for brightness profile estimation (bottom panel).
 }
  \label{lossfunctions}
\end{figure}

\begin{figure*}
 \centering
 \includegraphics[width=0.95\textwidth]{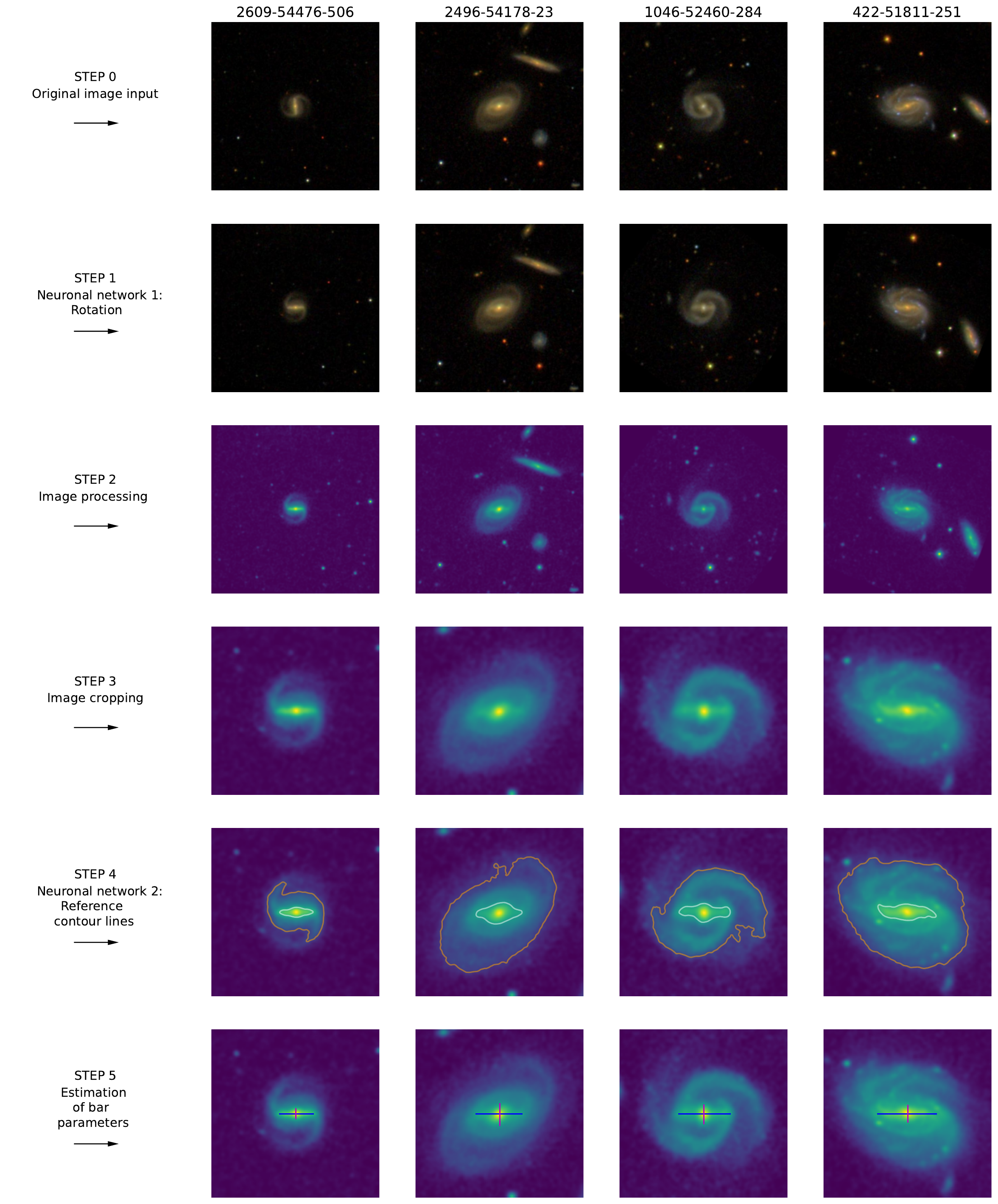}
\caption{Steps of the procedure used to calculate the bar parameters for four example objects from the analysis sample, which were not included in the neural network training processes.}
  \label{Muestra}
\end{figure*}


Galactic bars exhibit a wide variety of shapes and sizes, varying their strength, which affects their ability to channel the interstellar medium toward the central regions of the galaxies. Over time, bars undergo evolutionary changes: their pattern speeds decrease, causing them to become more elongated, eccentric, and consequently, stronger \citep{alth12}. In this context, bar structural characteristics can be accurately quantified through two-dimensional image modeling of the galaxy components.

In this work, we develop a method to estimate the main characteristic parameters of galactic bars, the semi-major axis ($a$) and the projected bar axis ratio ($(b/a)_{Bar}$). Since the analysis and control samples have the same number of objects, we started by randomly selecting $150$ objects from each ($300$ objects in total), and downloading the corresponding images using the Image Cutout tool from the SDSS web service\footnote{https://skyserver.sdss.org/dr16/en/help/docs/api.aspx\#imgcutout}. The images were downloaded with a predefined size of $500\times500$ pixels and a scale of $0.30$ arcsec/pixel, large enough to contain each galaxy with its disk.

For each image we visually identify the rotation angle of the bar in order to establish a new reference system, where the  $x$ direction extends along the bar and the $y$ direction is perpendicular to it. Subsequently, a series of filters were applied to facilitate the analysis of the brightness curves along the $x$ and $y$ axes: i) attenuation of the values in the B matrix of each RGB image, ii) application of a Gaussian blur with a $5\times5$ pixel kernel and a standard deviation of $30$ pixels, iii) conversion to grayscale using color coefficients suggested by the NTSC standard, and iv) normalization of brightness to values between $0$ and $1$.

After processing, we performed a visual inspection of each image and manually selected four reference points enclosing the bar: two on the $x$-axis and two on the $y$-axis. These points were identified in the image brightness profiles along each axis to obtain four brightness measurements. These values were averaged to derive a single iso-brightness contour in the image that encloses the bar and represents its two-dimensional profile. Fig. \ref{BarBrightness} illustrates the procedure: starting with the identification of the four reference points (A, B, C, and D), followed by their locations in the brightness distributions along the axes, and concluding with the estimation of the bar profile using the resulting contour.

Subsequently, the collected data were used to train two neural networks. The first, a ResNet50 convolutional neural network \citep{He2015} designed for processing RGB images and aligning them with the bar direction, was trained on an augmented dataset of $10000$ images. These images were generated by applying random rotations to the original $300$ images, cropped to $75\times75$ pixels, with $2000$ randomly selected for the validation sample. The training process used mean absolute error (MAE) as the loss function and Adam as the optimizer, with variable values for the learning rate, mini-batch size, and epochs. The upper panel of Fig. \ref{lossfunctions} shows the evolution of the loss function across epochs for both the training and validation sets, indicating a good convergence to approximately $2.1$ degrees.

The second neural network, a standard network with $8$ layers and $3800$ neurons, was trained from the brightness distributions of each image along the $x$ and $y$ axes, with the target defined as the average brightness value that constitutes the bar profile. This network was fed with $300$ vectors of $1000$ components containing the $x$ and $y$ brightness profiles of each image with the filters already applied.  During training, we used a cross-validation technique with $5$ different folds, each containing $60$ objects. Again, Adam was chosen as the optimizer and MAE as the loss function, with variable values for the learning rate, exponential decay rate, mini-batch size, and epochs. As shown in the lower panel of Fig. \ref{lossfunctions}, the fold-averaged loss functions for the training and validation sets across different epochs demonstrate rapid and effective convergence to approximately $0.025$.
Finally, we define the size of the galaxy disk as the greatest distance between two points on the iso-brightness contour of each image where the brightness has fallen to $15\%$ of its maximum value. To normalize the semi-major axis $a$, we divide its value by the characteristic distance of the galaxy disk. 

These procedures enabled us to estimate the semi-axes of the bars, $a$ and $b$, as the $x$-direction length and $y$-direction width of the bars reference contours, normalized to the disc scale, for all galaxies in our samples. Then, from these parameters we determined i) the ratio $b/a$ as well as ii) the physical size of the bars ($L_{Bar}$) in kpc using the images scale and the Hubble's law, iii) the bar size normalized to the galaxy disk size ($L_{Bar}$/\textsc{disk scale length}) and iv) the ellipticity of the bars ($\epsilon = 1 - (b/a)_{Bar}$).
Fig. \ref{Muestra} illustrates in an orderly manner the procedure followed to calculate the bar parameters for four objects from the analysis sample that were not used during the neural networks training processes. \\

\section{Effects of interactions in barred AGN galaxies}

\begin{figure}
\centering
\includegraphics[width=.43\textwidth]{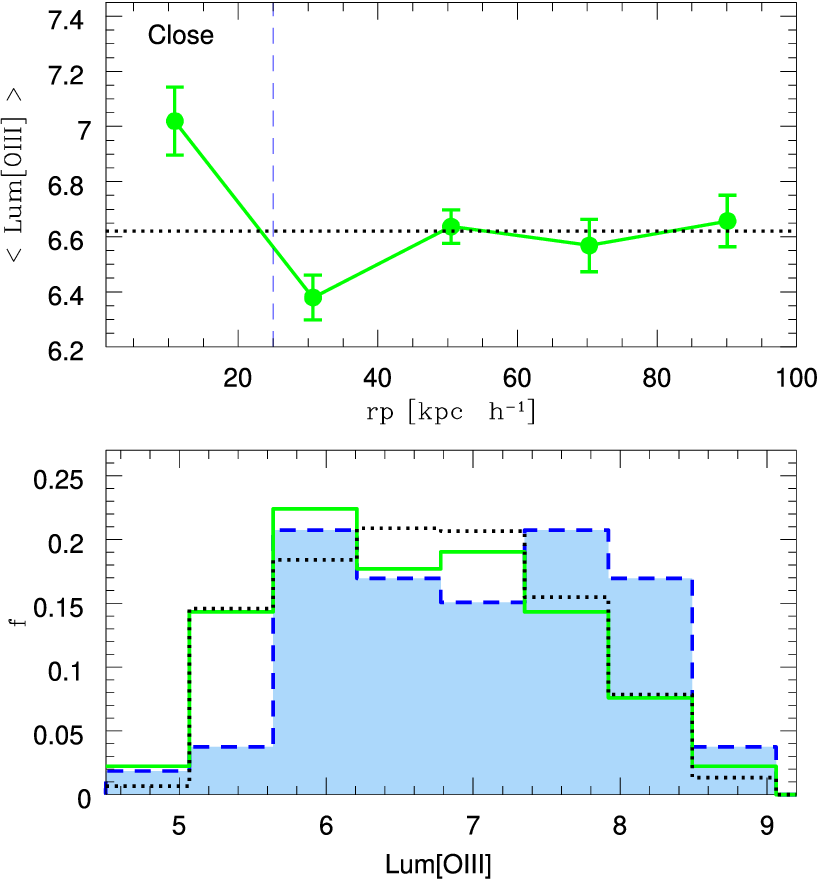}
\caption{Upper panel: < $Lum[OIII]$ > values as a function of projected distance, $r_p$, for barred AGN galaxies in pair systems (solid lines). The mean value of $Lum[OIII]$ for barred AGN galaxies in the control sample are plotted in horizontal dotted lines.
Lower panel: Distributions of $Lum[OIII]$ for barred AGN galaxies in pairs (solid lines), BGCP (dashed lines) and barred AGNs in the control sample (dotted lines).
}
 \label{LOrpH}
\end{figure}

\begin{figure*}
\sidecaption
  \includegraphics[width=12cm]{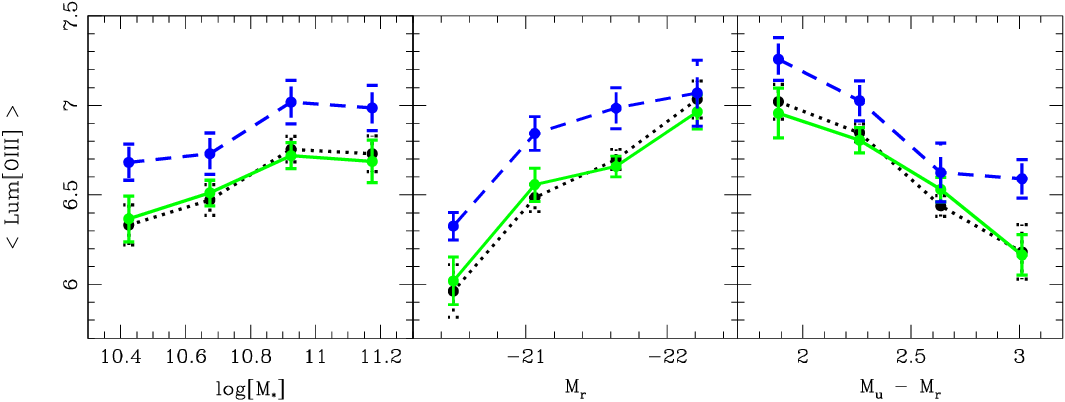}
     \caption{< $Lum[OIII]$ > as a function of stellar mass,  $log(M_*)$, $r-band$ magnitude, $M_r$, and color, $M_u - M_r$, for barred AGN galaxies in pair systems (solid lines), barred AGNs in close pairs (dashed lines) and barred AGN galaxies in the control sample (dotted lines).}
     \label{LOprop}
\end{figure*}

In this section, we conducted an analysis focusing on the combined influence of bars and interactions on the nuclear activity driven by the feeding of the central black hole.

To trace the nuclear activity of AGN galaxies, we focus on the dust-corrected luminosity of the [OIII]$\lambda$5007 line, denoted as $Lum[OIII]$. This particular emission line, [OIII], stands out as one of the most prominent narrow emission lines within optically obscured AGN, boasting minimal contamination from star formation within the host galaxy.
The effectiveness of the $Lum[OIII]$ estimator has been extensively explored by several authors through different analyses 
\citep{Kauffmann2003, heck04, heck05, Brinchmann2004}. Moreover, \cite{Kauffmann2003} discovered negligible contamination from star formation lines in $Lum[OIII]$ when examining host galaxies with high metallicity.
Within our dataset, the majority of AGN exhibit stellar masses $M^* >10^{10}$ $M_{\sun}$ (refer to  Fig. 2). This observation aligns with the mass-metallicity relation \citep{trem04}, suggesting high metallicities and consequently minimal contamination from star formation lines.

In Fig. \ref{LOrpH} (upper panel) we show the relation between the mean $Lum[OIII]$ as a function of projected distance of barred AGN galaxies in pair systems.
Errors were estimated by bootstrap resampling
techniques \citep{barrow} in this and the following figures.
The dotted horizontal line depict the mean $Lum[OIII]$ for the corresponding control sample. 
As can be seen, nuclear activity increases toward lower pair projected separations. However, for most pair separations, AGN activity is not enhanced relative to controls, with only barred AGN galaxies in very close pairs ($\approx$ $r_p <$ 25 kpc $h^{-1}$) showing significantly higher mean nuclear activity values compared to the control sample.
Moreover, we observe that the second bin ($r_p$ $\approx$ 30 kpc $h^{-1}$) shows a notable deficit in nuclear activity, with a decrease of $\approx$ 0.7 dex compared to the first bin. Additionally, this deficit is approximately 0.2 dex when compared to the bins with larger projected separations.
In the lower panel of Fig.~\ref{LOrpH}, the distributions of $Lum[OIII]$ are depicted for both barred AGN objects in pairs and barred active galaxies in the control sample. A clear similarity in the trend of both barred AGN samples is showed, displaying comparable distributions of nuclear activity. 
However, the most significant difference between the  $Lum[OIII]$ of barred AGN  in pairs and that of the control sample is obtained when the analysis is restricted to pairs with $r_p <$ 25 kpc $h^{-1}$. This subsample shows the strongest signal of enhanced nuclear activity. 
We can conclude that AGNs in pairs with small $r_p$ have a statistically significant increase of their  $Lum[OIII]$.
In consequence, we define a subsample of barred AGN galaxies in close pairs by imposing the restrictions, $r_p <$ 25 kpc $h^{-1}$; hereafter the BGCP sample.

\cite{Kauffmann2003} considered, from a sample of 22623 optically selected AGN using the BPT diagram, those classified as powerful AGN with $Lum[OIII] > 10^{7.0} L_{\odot}$. Their study also revealed that quasars (QSO) and strong type 2 AGNs, both meeting the criterion of $Lum[OIII] > 10^{7.0} L_{\odot}$, share similar characteristics associated with all types of AGN exhibiting strong $[OIII]$ emissions.
In the same direction, we defined in the present work, AGN galaxies with $Lum[OIII] > 10^{7.0} L_{\odot}$ as powerful active galaxies.
As can be clearly appreciated, barred AGNs in close pairs
(BGCP) show larger $Lum[O III]$ values than AGNs in the other samples. In fact, while 50$\%$ $\pm$ 0.57 of AGN in BGCP  sample have $Lum[O III]$ $>$ $10^7$ $L_{\odot}$, only 36$\%$ $\pm$ 0.40 and 38$\%$ $\pm$ 0.42 of barred AGN in pairs and in the control sample, respectively, show such a powerful activity.

In Fig.~\ref{LOprop}, we depict the mean $Lum[OIII]$ values relative to host galaxy characteristics for barred AGN galaxies within pair systems (solid lines), BGCP sample (dashed lines), and barred AGNs in the control sample (dotted lines).
The left panel illustrates the relationship between the mean $Lum[OIII]$ and the stellar mass of their respective host galaxies. Notably, higher nuclear activity is generally associated to more massive hosts. Additionally, barred galaxies within close pair systems exhibit an enhanced nuclear activity across all stellar mass ranges, compared to both the control sample and barred AGNs in pairs.
This suggests a distinct boost in nuclear activity for barred AGN inhabiting close pair systems, irrespective of the stellar mass of their host galaxies, compared to the other samples examined in this study.
In the same way, the following panel displays the mean $Lum[OIII]$ values relative to host luminosities ($M_r$), indicating that, overall, more luminous galaxies exhibit more efficient nuclear activity. Notably, barred AGNs within close pair systems demonstrate higher $Lum[OIII]$ levels compared to their counterparts in other samples across the entire range of $M_r$.

Additionally, the right panel in Fig.~\ref{LOprop} illustrates the mean $Lum[OIII]$ values, plotted against the $M_u-M_r$ color, for AGN galaxies in the different samples analyzed in this study.
From this analysis, a clear trend emerges, indicating an increase in nuclear activity with bluer colors across AGN galaxies in the different samples.  Besides, it's evident that the nuclear activity levels are comparable between barred AGN in pair systems and those in the control sample. However, barred active galaxies within close pair systems exhibit notably higher $Lum[OIII]$ values compared to the other two samples. 

The findings presented in this section indicate that external perturbations from a nearby galaxy companion can influence the dynamics of gas flows induced by galactic bars, leading to an additional increase in central nuclear activity in barred AGN galaxies located in pair systems.
In this context, the coexistence of both perturbations, galactic bars and interactions significantly amplifies the central nuclear activity. This effect is particularly pronounced in host galaxies that are luminous, massive and exhibit blue colors.

\section{Relation between AGN activity and  bar strength parameters}

Considering the bar strength is essential when examining the influence of bars on galactic dynamics \citep{cist13}. 
The non-axisymmetric potential of a bar can significantly impact the overall potential of a galaxy \citep{sell93,Mart95,Binn09}. This impact is contingent upon the bar strength, which can be defined in terms of its ability to exert torques on the interstellar medium due to its gravitational potential.
In this regard, \cite{zee23} propose that the impact of bar-driven evolution on star formation and black hole activity should be assessed using various bar parameters, including bar size, $L_{Bar}$, and the axial ratio, $(b/a)_{Bar}$. To explore the correlation between bar strength and AGN activity, as well as the role of interactions, in this section we analyze different indicators of bar strength and their relationship with central nuclear activity.

\begin{figure}
 \centering
 \includegraphics[width=.5\textwidth]{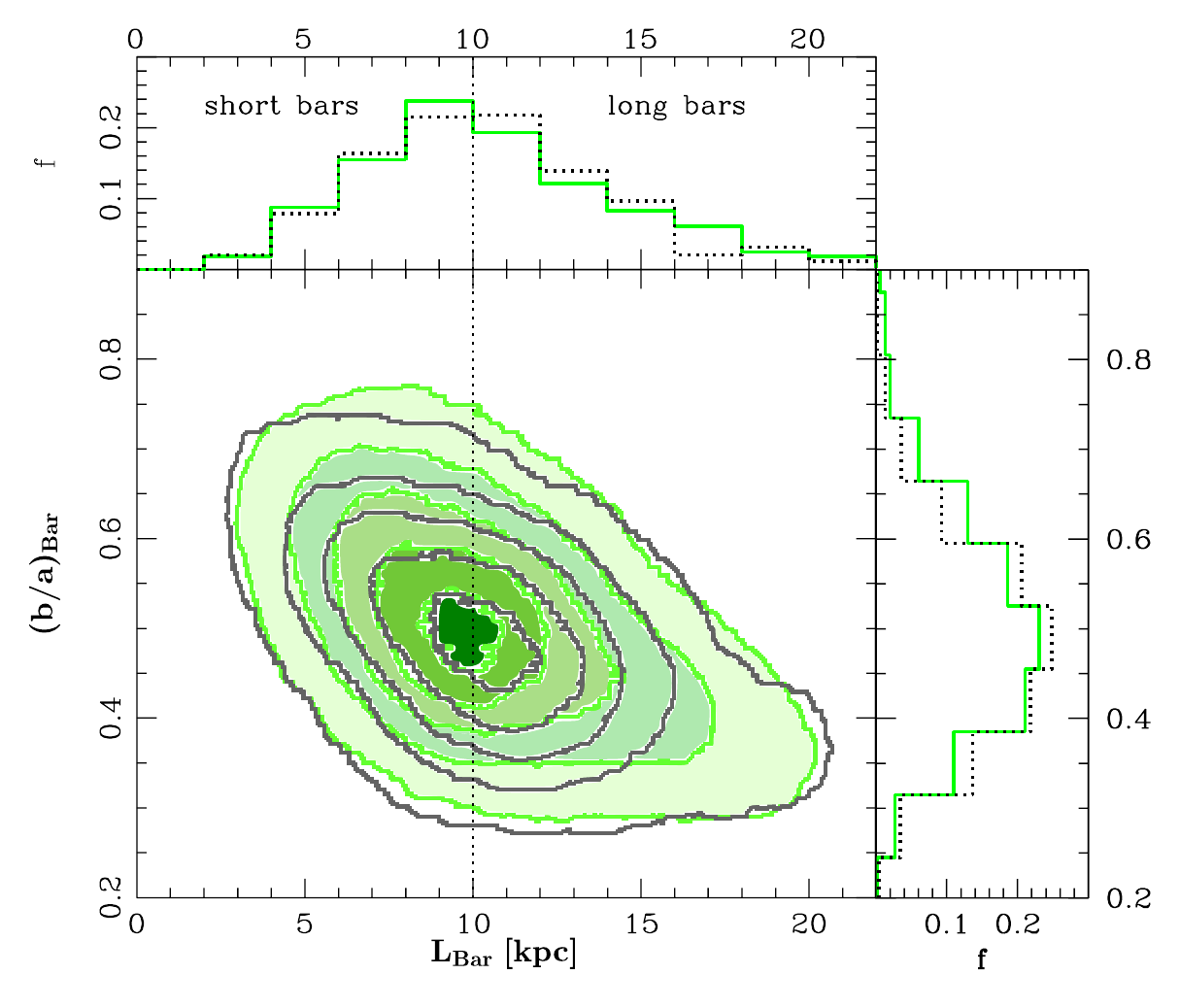}
\caption{Projected bar axis ratio, $(b/a)_{Bar}$ as a function of physical size of bars, $L_{Bar}$, for barred AGN galaxies in pair systems (represented by green contours) and barred AGNs in the control sample (represented by gray contours). 
The upper and right panels illustrate the normalized distributions of $L_{Bar}$ and $(b/a)_{Bar}$, respectively, for each sample.
The dotted vertical black line marks the mean value of $L_{Bar}$, used to divide the sample into two categories, galaxies with short and long bars.
 }
  \label{baLbar}
\end{figure}

In Fig.~\ref{baLbar}, we observe the projected bar axis ratio, $(b/a)_{Bar}$ as a function of the physical size of bars, $L_{Bar}$, among barred AGN galaxies, comparing those within pair systems and their counterparts from the control sample. Notably, a consistent trend emerges across both samples showing that longer bars tend to exhibit greater elongation. This observation aligns well with findings from previous studies \citep{buta15,conso16,lee20,zee23}.
Additionally, the distributions of $L_{Bar}$ and $(b/a)_{Bar}$ for both barred AGN samples  display similar trends, as depicted in the top and right panels, respectively. 
The dotted vertical line in Fig.~\ref{baLbar} represents the mean value of $L_{Bar}$ $\approx$ 10 kpc $h^{-1}$, which we consider as the appropriate threshold for categorizing the samples into galaxies with long and short bars.

\begin{figure}
 \centering
 \includegraphics[width=.37\textwidth]{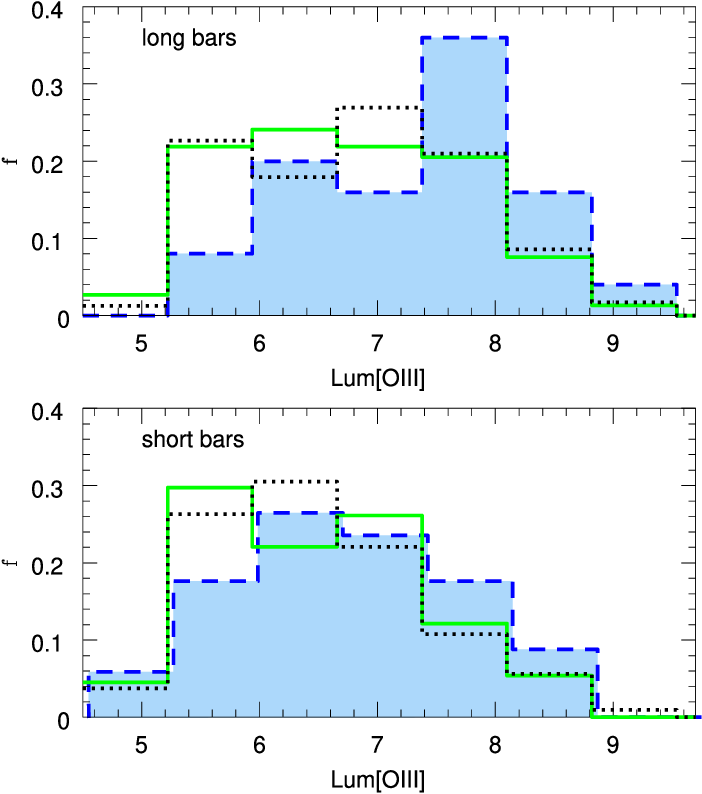}
\caption{Normalized distributions of $Lum[OIII]$ for barred AGN galaxies in pair systems (solid lines), barred AGNs in close pairs (dashed lines) and their counterparts in the control sample (dotted lines), taking into account long and short bars (upper and bottom panels, respectively).
 }
  \label{hLObarls}
\end{figure}

Furthermore, in Fig.~\ref{hLObarls}, we observe the distributions of $Lum[OIII]$ among barred AGN objects in pairs, the BGCP sample, and barred active galaxies in the control sample, considering both, the short and long bars sub--samples. It is noteworthy that barred AGN galaxies with long barred structures exhibit a more pronounced signal of enhanced nuclear activity compared to those with short bars, across the different samples examined in this study. A distinct similarity is evident in the trends among the different barred AGN samples, reflecting comparable distributions of nuclear activity. However, a notable deviation arises in barred AGN galaxies located within close pair systems, where a significant surplus of high $Lum[OIII]$ values is evident compared to the other samples. This trend is particularly pronounced in galaxies with long bars. Table \ref{tab:bar} quantifies these trend given the percentage of AGN galaxies with long and short barred structures that exhibit powerful nuclear activity ($Lum[OIII] > 10^{7.0} L_{\odot}$) for the different samples under study.

\begin{table}
\center
\caption{Percentages of AGN galaxies with long and short bars that present powerful nuclear activity values in the samples studied in this work. Standard errors are also included.}
\begin{tabular}{|c c c | }
\hline
Restriction &\multicolumn{2}{c|}{$Lum[OIII] > 10^{7.0} L_{\odot}$}   \\
\hline
Samples & $\%$ of short bars & $\%$ of long bars \\
\hline
\hline
Control  &   27.7$\%$ $\pm$ 0.43 &   45.7$\%$ $\pm$ 0.51 \\
Pairs  &    28.2$\%$ $\pm$ 0.37 & 42.4$\%$ $\pm$ 0.46 \\
BGCP  &    32.2$\%$ $\pm$0.65 & 67.4$\%$ $\pm$ 0.72 \\
\hline
\end{tabular}
{\small}
\label{tab:bar}
\end{table}

\begin{figure*}
\sidecaption
  \includegraphics[width=12cm]{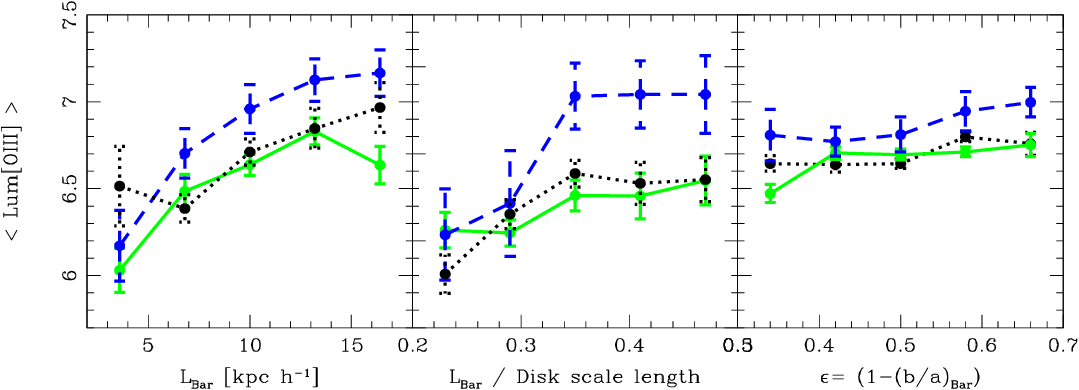}
\caption{< $Lum[OIII]$ > as a function of the physical size of bars, $L_{Bar}$, (left panel), bar size normalized to the disk scale length, $L_{Bar}$/\textsc{Disk scale length} (middle panel) and bar ellipticity, $\epsilon$ = 1-$(b/a)_{Bar}$ (right panel),  for barred AGN galaxies in pair systems (solid lines), barred AGNs in close pairs (dashed lines) and barred AGN galaxies in the control sample (dotted lines).}
  \label{Lobar}
\end{figure*}

To explore the direct correlation between bar features, AGN activity and the impact of galactic interactions, we compare the nuclear activity indicator with bar strength measures in the galaxies from our samples, in the following analysis.
 Fig. \ref{Lobar} shows the relationship between the mean values of $Lum[OIII]$ and two bar size metrics: the physical size of bars ($L_{Bar}$, left panel) and the bar size normalized to the galaxy disk size ($L_{Bar}$/\textsc{disk scale length}, middle panel) for barred AGN objects in pairs, BGCP sample and barred active galaxies in the control sample. The data indicates that galaxies with longer bars tend to exhibit more efficient nuclear activity. Notably, barred AGNs within close pair systems consistently show higher $Lum[OIII]$ levels compared to their counterparts in other samples, across the entire range of $L_{Bar}$ and $L_{Bar}$/\textsc{disk scale length}.

Galactic bars exhibit axis ratios ranging from oval distortions to highly elongated and well-defined structures. According to \cite{Mart95}, the axis ratio of a bar can serve as a measure of its strength, as it directly relates to the non-axisymmetry of the bar's potential. 
Consequently, the ellipticity ($\epsilon = 1 - (b/a)_{Bar}$) can effectively quantify bar strength.
In this context, the right panel of Fig.  \ref{Lobar} shows the mean $Lum[OIII]$ values as a function of $\epsilon$, revealing a very slight and subtle trend: galaxies with greater nuclear activity exhibit greater elongation. However, considering the errorbars, no clear correlation is observed between central nuclear activity and bar elongation, for the different samples analyzed in this work. 
\cite{silva22} investigated the role of bars on AGN feeding using a sample of galaxies based on SDSS DR2, suggesting that the galaxies with the most prominent bars tend to have higher activity levels on average, however given the scatter in the data this result is not conclusive.

\section{Major and Minor pairs}

\begin{figure}
 \centering
 \includegraphics[width=.47\textwidth]{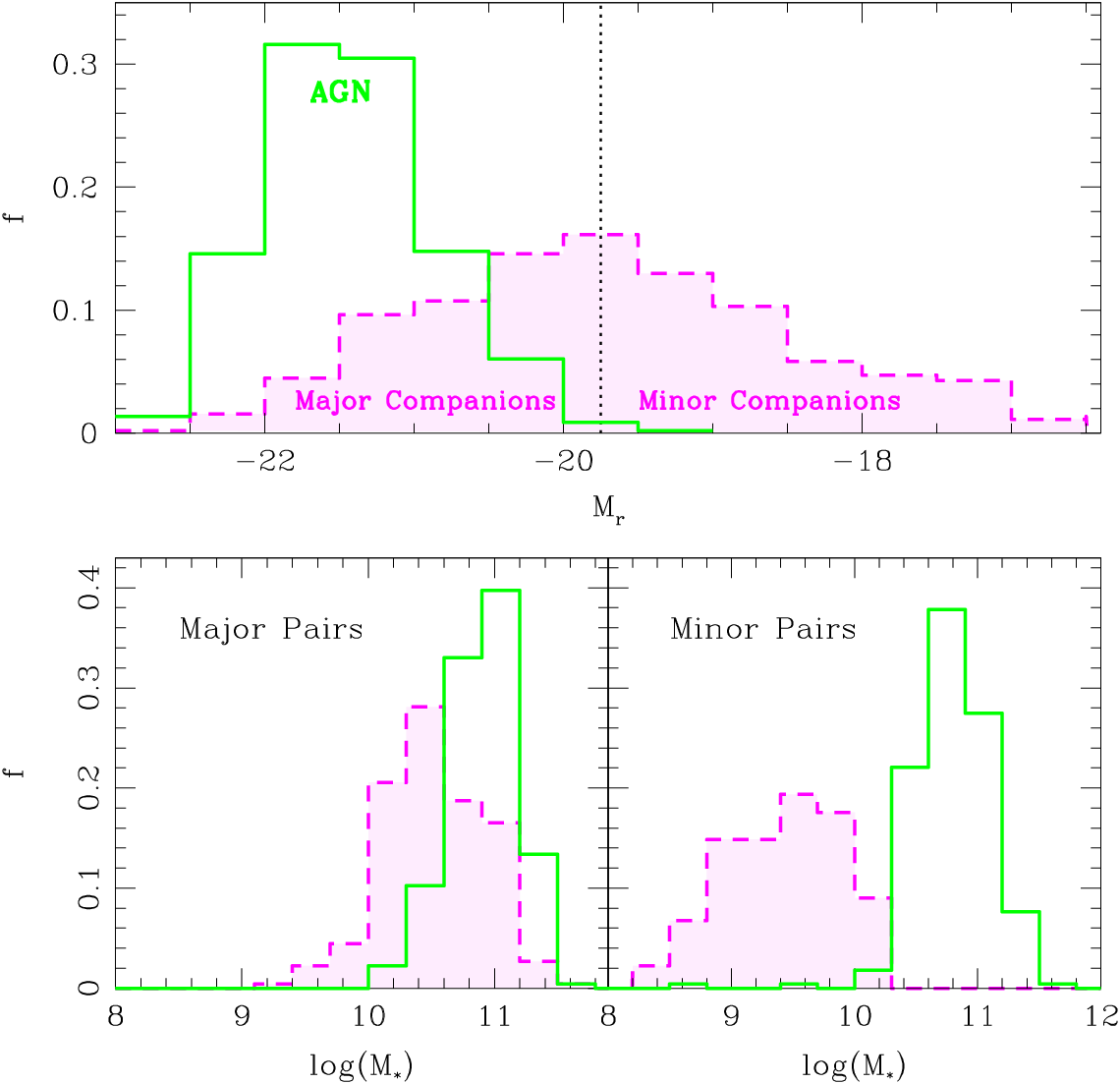}
\caption{Upper Panel: Normalized distributions of luminosity ($M_r$) for barred AGN galaxies (solid line) and their galaxy companions (dashed line). The dotted vertical line represents the mean value of the distribution.
Lower Panels: Normalized distributions of stellar masses ($log(M_*)$) for barred AGN galaxies (solid lines) and their galaxy companions (dashed lines) in major and minor pairs (left and right panels, respectively).
 }
  \label{histL}
\end{figure}

\begin{figure*}
\sidecaption
  \includegraphics[width=12cm]{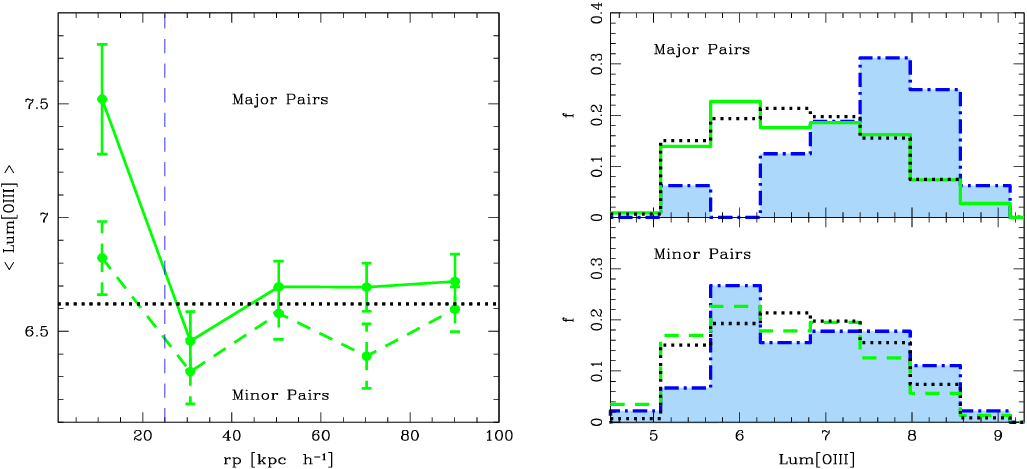}
\caption{Left panel: < $Lum[OIII])$ > as a function of projected distance, $r_p$, for barred AGN galaxies in major and minor pairs (solid and dashed lines, respectively). The mean value of $Lum[OIII]$ for barred AGN galaxies in the control sample is plotted with horizontal dotted line.
Right panels: Normalized distributions of $Lum[OIII]$ for barred AGN galaxies (solid lines) and BGCP (dot-dashed lines) in major and minor pair systems (upper and lower panels, respectively). Barred AGN galaxies in the control sample are plotted with dotted lines.
 }
  \label{LrpMyM}
\end{figure*}

\begin{figure*}
\sidecaption
  \includegraphics[width=12cm]{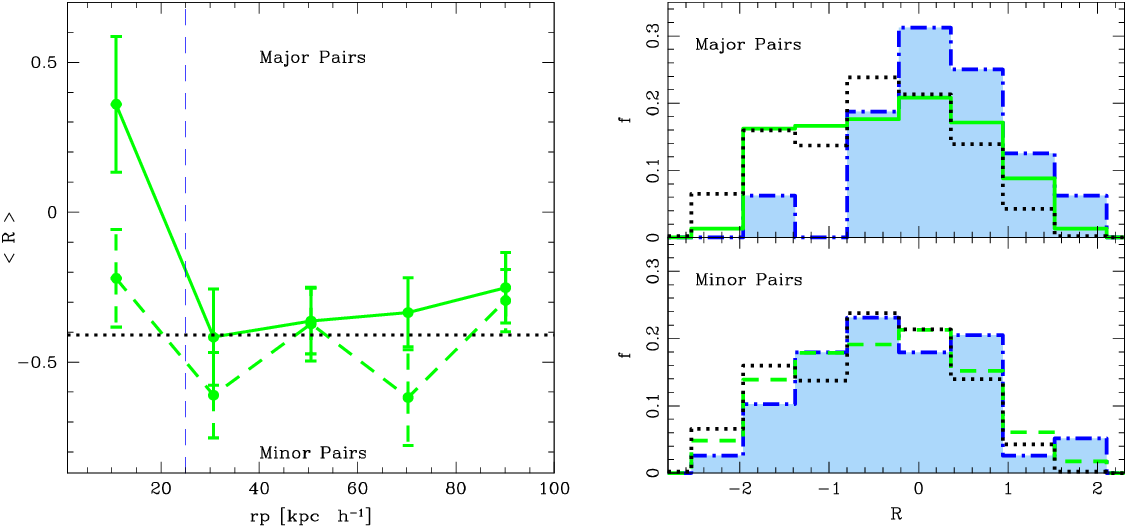}
\caption{Left panel: < $\cal R$ > as a function of projected distance, $r_p$, for barred AGN galaxies in major and minor pairs (solid and dashed lines, respectively). The mean value of $\cal R$ for barred AGN galaxies in the control sample is plotted with horizontal dotted lines.
Right panels: Normalized distributions of $\cal R$ for barred AGN galaxies (solid lines) and BGCP (dot-dashed lines) in major and minor pair systems (upper and lower panels, respectively).
Barred AGN galaxies in the control sample are plotted with dotted lines.
 }
  \label{RrpMyM}
\end{figure*}

The effects of companion galaxies depend significantly on the relative luminosity relationship between the member galaxies in pair systems. 
Previous studies have explored major and minor interactions by considering various galaxy features. In this sense, several authors (e.g. \citealt{Lambas2003, Lambas2012}, \citealt{Ellison2008, Ellison2010, das21}) indicate that interactions within galaxy pairs can enhance star formation activity, particularly in the brighter pair member and in major mergers. 
In the realm of AGN studies, \cite{Alonso2007} examined active galaxies within close pair systems, distinguishing between those with a bright companion and those with a faint companion. 
Their research revealed that AGN hosts with bright companions exhibited a significant increase in $Lum[OIII]$ compared to those with faint companions.
\cite{capelo15} used high-resolution hydrodynamical simulations to demonstrate that black hole accretion in merging galaxies is primarily influenced by the initial mass ratio. Their findings indicate that major mergers tend to increase the disparity between black hole masses, while minor mergers lead to a reduction in mass contrast.
They also found that both merger-related and secular processes contribute to AGN activity.
Additionally, A18 observed a significant dependency of nuclear activity on the properties of galaxy pair companions in AGN residing within pair systems. They showed that nuclear activity in pairs experiences a notable increase when galaxy companions exhibit brighter and more massive hosts, higher metallicities, bluer colors, efficient star formation activity, and young stellar populations.
More recently, \cite{micic24} found that interactions between low-mass galaxies can significantly trigger black hole activity, providing new insights into the growth and evolution of black holes in the early universe.

In this section, we explore how major and minor interactions influence the central nuclear activity in barred AGN galaxies within pair systems. This study aims to enhance our understanding of a topic that has been sparsely investigated in the context of barred galaxies.
For the current analysis, we have categorized our sample into major and minor pairs based on variations in the luminosity of the galaxy pair companions. Specifically, we are looking at barred AGN galaxies with both bright and faint companions, using the mean value of the absolute magnitude distribution, $M_r$ = -19.75, as the dividing threshold. Normalized distributions of luminosity for barred AGN galaxies and their associated galaxy companions are shown in the upper panel of Fig.~\ref{histL}. 
Additionally, the lower panels exhibit distributions of stellar masses for barred AGN galaxies and their companions, clearly showing both similar and distinct $log(M_*)$ distributions within major and minor pairs, respectively.

In Fig. \ref{LrpMyM} (left panel), the relationship between the mean $Lum[OIII]$ and the projected distance of barred AGN galaxies within major and minor pair systems is depicted. The dotted horizontal line represents the mean $Lum[OIII]$ values for the corresponding control sample. 
AGN activity is generally not enhanced relative to the control sample for most separations; however, there is a notable escalation in nuclear activity as the pair projected separations decrease.
This pattern is particularly pronounced in major systems, whereas minor pairs do not exhibit a noticeable increase in nuclear activity compared to galaxies in the control sample.
Moreover, in major systems, we observe that the second bin ($r_p$ $\approx$ 30 kpc $h^{-1}$) exhibits a notable deficit in nuclear activity, similar to what is observed in Fig. \ref{LOrpH}. This deficit is approximately 1 dex lower than in the first bin and around 0.25 dex lower compared to bins with larger projected separations.
In the right panel of Fig.~\ref{LrpMyM}, the distributions of $Lum[OIII]$ are illustrated for barred AGN objects within major and minor pairs, alongside barred active galaxies from the control sample. Both barred AGN samples exhibit a notable similarity in their trends, showcasing comparable distributions of nuclear activity within both major and minor pairs. However, a clear  deviation emerges among barred AGN galaxies located in close pair systems. Here, a significant surplus of high $Lum[OIII]$ values is evident in major pairs compared to the other samples.

Furthermore, with the goal of analyzing the accretion strength of central black holes in barred AGN galaxies located in major and minor pair systems, we have computed the accretion rate parameter, $\mathcal{R} = \log(Lum[OIII] / M_{BH})$ \citep{heck04}. Utilizing the correlation between black hole mass, $M_{BH}$, and bulge velocity dispersion, $\sigma_*$ \cite{trem04}, we initially estimated $M_{BH}$ for the samples examined in this study:

\begin{equation}
\log M_{BH} = \alpha + \beta \log(\sigma_* / 200).
\end{equation}

According to \cite{gra08}, the central velocity dispersion is influenced by stellar motion along a bar, consequently altering the $M_{BH} - \sigma_*$ relation in barred galaxies. We adopted ($\alpha$, $\beta$) = (7.67 $\pm$ 0.115, 4.08 $\pm$ 0.751) for barred active galaxies in our samples \citep{gult09}. For this analysis, we constrained $\sigma_* > 70 km s^{-1}$, given that the instrumental resolution of SDSS spectra is $\sigma_* \approx$ 60 to 70 $km s^{-1}$.

To supplement the preceding analysis, the left panel of Fig. \ref{RrpMyM} illustrates the $\cal{R}$ parameter plotted against $r_p$ for barred AGN galaxies in major and minor pairs and for galaxies in the control sample (represented by the dotted horizontal line). 
The accretion rate is generally not enhanced relative to the control sample for most separations, although there is a notable increase as the pair projected separations decrease.
This trend is especially pronounced in major systems, whereas minor pairs show no significant rise in accretion strength compared to galaxies in the control sample.
In addition, the right panels display the distributions of $\cal{R}$ for barred AGN galaxies in pair systems, including close pairs, and in the control sample for both major and minor systems. 
Both samples of barred AGNs display a notable similarity in their patterns, with comparable distributions of accretion rate values for barred AGNs in both paired systems and their  counterparts from the control sample. This trend is evident in both major and minor pair systems. Additionally, barred AGNs in major pairs typically exhibit higher $\cal{R}$ values compared to those in minor systems.
However, a notable deviation occurs among barred AGN galaxies located in major pair systems with $r_p <$ 25 kpc $h^{-1}$. In this context, there is a significant surplus of high $\cal{R}$ values observed in barred AGN galaxies within major pairs with close companions, in contrast to the other samples.
These trends are quantified in Table \ref{tab:barMM} that shows the percentage of barred AGN galaxies in minor and major pairs that exhibit powerful nuclear activity ($Lum[OIII] > 10^{7.0} L_{\odot}$) and efficient accretion rate ($\cal R$ $>$0), for the different samples studied in this work.

\begin{table}
\center
\caption{Percentages of barred AGN galaxies in minor and major pairs that present powerful nuclear activity and high accretion rate values in the sample studied in this work. Standard errors are also included.}
\begin{tabular}{|c c c| }
\hline
Samples &  Minor Pairs & Major Pairs \\
\hline
\hline
Restriction &\multicolumn{2}{c|}{$Lum[OIII] > 10^{7.0} L_{\odot}$}   \\
\hline
Pairs  &    36.5$\%$ $\pm$ 0.39 & 39.9$\%$ $\pm$ 0.45 \\
BGCP  &    37.3$\%$ $\pm$0.43 & 73.7$\%$ $\pm$ 0.56 \\
Control &\multicolumn{2}{c|} {38.0$\%$ $\pm$0.42}  \\
\hline
Restriction &\multicolumn{2}{c|}{$\cal R$ $>$0}   \\
\hline
Pairs  &    33.1$\%$ $\pm$ 0.38 & 39.8$\%$ $\pm$ 0.43 \\
BGCP  &    35.8$\%$ $\pm$0.52 & 64.7$\%$ $\pm$ 0.68 \\
Control &\multicolumn{2}{c|} {32.8$\%$ $\pm$0.39} \\
\hline
\hline
\end{tabular}
{\small}
\label{tab:barMM}
\end{table}

These findings reveal that the movement of material generated by galactic bars toward the central regions of active galaxies, where supermassive black holes reside, is significantly enhanced and heavily influenced by the proximity of a nearby massive galaxy companion. This neighboring galaxy exerts gravitational forces that can intensify the inward transport of gas and dust, effectively fueling the central black hole and potentially increasing its accretion rate and nuclear activity.


\section{Summary and conclusions}

We analyzed the impact of mergers and interactions on the central nuclear activity of barred spiral AGN galaxies. Our study focuses on a sample of barred active galaxies in paired systems with projected separations of less than 100 kpc $h^{-1}$ and relative radial velocities of less than 500 km $\rm s^{-1}$, all within a redshift range of z$<$0.1.

To accurately quantify how interactions influence the material transport by galactic bars to central black holes and the resulting nuclear activity, we also constructed a control sample of barred spiral AGN galaxies lacking paired companions. Moreover these galaxies were selected to match the paired sample in redshift, absolute $r$-band magnitude, stellar mass, color, and stellar age distributions.

Furthermore, galactic bars display a wide range of shapes and sizes, which may influence their ability to channel the material toward the central regions of galaxies. 
To explore this, we used machine learned inference models to calculate the structural characteristics of the bars through two-dimensional image modeling of the galaxy components, and examined their relationship with central nuclear activity and the role played by galactic interactions.

The main results and conclusions of our analysis are summarized as follows:

(i) We found that nuclear activity increases as the projected separations of galaxy pairs decrease. However, it is important to emphasize that for most pair separations, AGN activity is not enhanced relative to isolated galaxies in the control sample. Only barred AGN galaxies in very close pairs ($\approx$ $r_p <$ 25 kpc $h^{-1}$) exhibit significantly higher nuclear activity compared to isolated galaxies in the control sample.
In this direction, we observed that while 50$\%$ $\pm$ 0.57 of barred AGN in the close pair sample have $Lum[O III]\ >\ 10^7$ $L_{\odot}$, only 36$\%$ $\pm$ 0.40 and 38$\%$ $\pm$ 0.42 of barred AGN in pairs and the control sample, respectively, exhibit such potent nuclear activity.

(ii) We also analyzed the mean $Lum[OIII]$ values in relation to the  host galaxy properties. Our findings indicate that higher nuclear activity is generally associated with more luminous, massive, and bluer host galaxies.
While nuclear activity levels are similar between barred AGN in pair systems and those in the control sample, barred galaxies with close pair companion exhibit enhanced nuclear activity across all ranges of luminosity, stellar mass, and color, compared to both the control sample and the overall barred AGNs in pairs.

(iii) From the measured values of bar physical size ($L_{Bar}$) and projected bar axis ratio ($(b/a)_{Bar}$), a consistent trend is observed: longer bars tend to be more elongated. This tendency is evident in both samples, including barred AGN galaxies in pair systems and their counterparts from the control sample.

(iv) We identified that barred AGN galaxies with long bar structures exhibit a more efficient nuclear activity compared to those with short lengths. A clear similarity is evident in the trends among the barred AGN galaxies in pairs and in the control sample, showing comparable distributions of nuclear activity. However, a notable deviation arises in barred AGN galaxies located within close pair systems, where a significant surplus of high $Lum[OIII]$ values is evident compared to the other samples. This trend is more significant in galaxies that present long bars. 

(v) To analyze the correlation between bar properties and AGN activity, and to understand the role of interactions, we compared  the AGN nuclear activity indicator with measures of bar strength. The study showed that galaxies with longer bars tend to exhibit more efficient nuclear activity. Notably, barred AGNs within close pairs consistently present higher $Lum[OIII]$ levels with respect to their counterparts in other samples, across the entire range of $L_{Bar}$ and $L_{Bar}$/\textsc{disk scale length}.
However, no clear correlation is observed between central nuclear activity and bar elongation across the different samples analyzed in this study.

(vi) We also examine the central nuclear activity in barred AGN galaxies undergoing major and minor interactions, considering the variations in the luminosity of the pair galaxy companions. 
We found a clear escalation in nuclear activity as the pair projected separations decrease, a pattern particularly pronounced in major systems. 
However, for most pair separations, AGN activity is not significantly enhanced compared to isolated galaxies in the control sample, with the notable increase occurring only in very close pairs.
In contrast, minor pairs do not exhibit a noticeable increase in nuclear activity compared to galaxies in the control sample.
Additionally, nuclear activity distributions in barred AGN samples within major and minor pairs exhibit similar trends. However, a notable deviation occurs among barred AGN galaxies in close pair systems, within  major pairs showing a significant surplus of high $Lum[OIII]$ values.

(vii) We also examined the accretion strength onto central black holes for barred AGN galaxies across various samples. 
The accretion rate is generally not elevated compared to the control sample for most pair separations. However, a noticeable increase in the accretion rate is observed as projected separations decrease, with this effect being particularly pronounced in major systems.
Minor pairs, however, show no significant rise in accretion strength compared to galaxies in the control sample. Furthermore, our analysis of the $\cal{R}$ distributions indicates that barred AGN galaxies within major pairs with close companions exhibit an excess of high accretion rate values compared to other samples.

This study reveals that in active galactic nuclei galaxies, the simultaneous coexistence of both phenomena, galactic bars and galaxy interactions, significantly boosts the central nuclear activity of supermassive black holes.
Specifically, the results indicate that the external gravitational influence from galaxy interactions disturbs the galactic material and enhances the efficacy of the bar structure in funneling this material toward the core of the galaxies. This dual mechanism interaction-induced disturbance and bar-driven inflow leads to a more efficient transport of gas and dust to the central regions, thereby increasing the accretion rate onto the supermassive black hole. Consequently, this results in a marked escalation in nuclear activity. The finding suggests that the synergy of galactic bars and interactions plays a crucial role in amplifying the central nuclear activity, significantly impacting the accretion processes and overall dynamics within AGN galaxies.

\begin{acknowledgements}
      This work was partially supported by the Consejo Nacional de Investigaciones
Cient\'{\i}ficas y T\'ecnicas and the Secretar\'{\i}a de Ciencia y T\'ecnica 
de la Universidad Nacional de San Juan. V.M. also acknowledges  support  from  project  DIDULS Regular N° PR2353857.

Funding for the SDSS has been provided by the Alfred P. Sloan
Foundation, the Participating Institutions, the National Science Foundation,
the U.S. Department of Energy, the National Aeronautics and Space
Administration, the Japanese Monbukagakusho, the Max Planck Society, and the
Higher Education Funding Council for England. The SDSS Web Site is
http://www.sdss.org/.

The SDSS is managed by the Astrophysical Research Consortium for the
Participating Institutions. The Participating Institutions are the American
Museum of Natural History, Astrophysical Institute Potsdam, University of
Basel, University of Cambridge, Case Western Reserve University,
University of
Chicago, Drexel University, Fermilab, the Institute for Advanced Study, the
Japan Participation Group, Johns Hopkins University, the Joint Institute for
Nuclear Astrophysics, the Kavli Institute for Particle Astrophysics and
Cosmology, the Korean Scientist Group, the Chinese Academy of Sciences
(LAMOST), Los Alamos National Laboratory, the Max-Planck-Institute for
Astronomy (MPIA), the Max-Planck-Institute for Astrophysics (MPA), New Mexico
State University, Ohio State University, University of Pittsburgh, University
of Portsmouth, Princeton University, the United States Naval Observatory, and
the University of Washington.
\end{acknowledgements}


%
   \bibliographystyle{aa} 
  \bibliography{biblio} 
%

\end{document}